\documentclass[%
twocolumn,
showpacs
 amsmath,amssymb,
prl,
]{revtex4-1}

\usepackage{%
 graphicx,
 hyperref,
 dcolumn,
}

\newcommand{\e}[1]{
\left[ #1 \right]
}

\newcommand{\abs}[1]{
\left|#1\right|
}

\newcommand{\rk}[1]{
\left( #1\right)
}

\newcommand{\rem}[1]{}

\newcommand{\ket}[1]{\left| #1 \right>}

\newcommand{\tn}[1]{\textnormal{#1}}

\begin{document}

\title{Klein-Tunneling of a Quasirelativistic Bose-Einstein Condensate in an Optical Lattice}

\author{Tobias Salger}
\email{salger@iap.uni-bonn.de}
\author{Christopher Grossert}
\email{grossert@iap.uni-bonn.de}
\author{Sebastian Kling}
\author{Martin Weitz}

\affiliation{Institut f\"ur Angewandte Physik der Universit\"at Bonn,
Wegelerstr. 8, 53115 Bonn}

\date{\today}

\begin{abstract}

Optical lattices have proven to be powerful systems for quantum simulations of solid state physics effects. Here we report a proof-of-principle experiment simulating effects predicted by relativistic wave equations with ultracold atoms in a bichromatic optical lattice that allows for a tailoring of the dispersion relation. We observe the analog of Klein-tunneling, the penetration of relativistic particles through a potential barrier without the exponential damping that is characteristic for nonrelativistic quantum tunneling. Both linear (relativistic) and quadratic (nonrelativistic) dispersion relations are investigated, and significant barrier transmission is observed only for the relativistic case. 
\end{abstract}

\pacs{??.??.??, ??.??.??, ??.??.??}

\maketitle

Klein tunneling, a textbook effect in which relativistic particles penetrate through a potential barrier without the exponential damping that is characteristic for nonrelativistic quantum tunnelling \cite{Klein29,Bjork64}, has never been observed for elementary particles. In this counterintuitive consequence of relativistic quantum mechanics, a strong potential, being repulsive for particles and attractive for antiparticles, results in particle- and antiparticle-like states aligning in energy across the barrier. Therefore, a high transmission probability is expected when a potential drop of the order of the particle’s rest energy $mc^2$ is achieved over the Compton length $h/mc$. For electrons, one derives an extremely high required electric field strength of $\approx 10^{16}$V/cm, which so far has prevented an experimental realization on this system. The observation of Klein tunnelling has however been reported in solid state analogons \cite{Kats06,Young09,Stand09,Steele09}, for example in graphene material. In this carbon material owing to a linear, i.e. quasirelativistic, dispersion relation around the Fermi edge, relativistic effects can be very illustratively emulated \cite{Kats06}. Other systems suitable for the simulation of quasirelativistic effects are ions with a long-lived two-component electronic structure in Paul traps \cite{Gerrit10}, where the nonrelativistically forbidden entry into a high potential well has been observed \cite{Gerrit11}. Experiments in photonic structures \cite{Longhi10} and in dark state media \cite{Juze08,Vais08} have been proposed. Ultracold atoms in optical lattices \cite{Bloch08} allow for the investigation of both linear and nonlinear Hamiltonians due to the neutral charge of the atoms, with prospects including the simulation of interacting relativistic quantum field theories \cite{Cirac10}.

Here we report a proof-of-principle quantum simulation of relativistic wave equation predictions with ultracold atoms in an optical lattice. Our experiment is based on rubidium atoms in a Fourier-synthesized lattice potential consisting of an optical standing wave with spatial periodicity $\lambda/2$, where $\lambda$ denotes the laser wavelength, and a higher spatial harmonic with $\lambda/4$ spatial periodicity. For a suitable choice of relative phases and amplitudes of the harmonics, the dispersion relation in the region between the first and second excited Bloch band becomes linear, as known for ultrarelativistic particles. We experimentally demonstrate both the transmission of atoms through a potential barrier for the case of a linear dispersion relation, i.e. Klein-tunneling, and the usual reflection of atoms by the barrier for the case of a quadratic, i.e. Schr\"odinger-like, dispersion in an excited Bloch band.

\begin{figure}
\includegraphics[width=8cm]{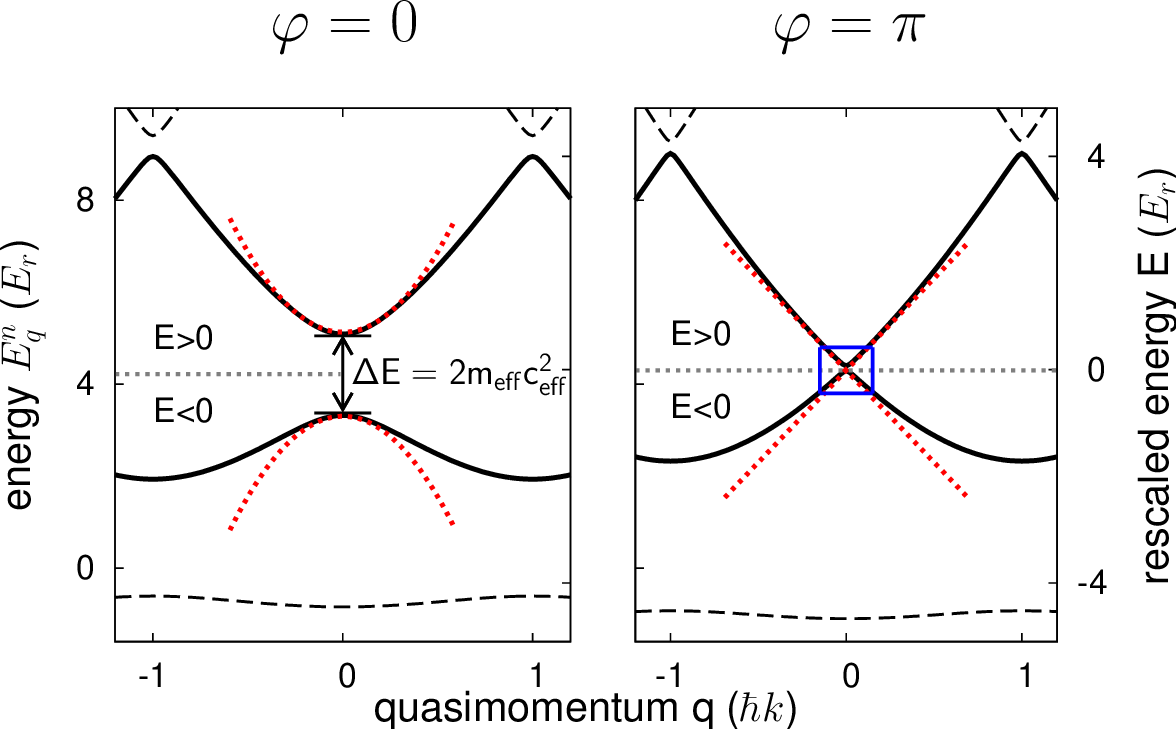}
\caption{Band structure for an optical lattice potential  for a relative phase of $\varphi=0$ (left) and  $\varphi=\pi$ (right). The parameters for the lattice depths were $V_1=5E_r$ and $V_2=1.6E_r$. For $\varphi=\pi$ the splitting between the first and the second excited band vanishes, and a Dirac-point (marked by the solid blue box) is observed. The bands relevant for the experiment are shown by solid lines, and on the rescaled energy scale shown on the right hand side, zero energy is chosen at the position of the band crossing.}
\end{figure}

The periodic potential used to taylor the dispersion of ultracold rubidium atoms is of the form $V\rk{z}=V_1/2\cos\rk{2kz} + V_2/2\cos\rk{4kz+\varphi}$, where $k=2\pi/\lambda$ is the photon wavevector, $V_1$ denotes the potential depth of the usual standing wave potential and $V_2$ that of the higher spatial harmonic, generated by the dispersion of multiphoton Raman transitions \cite{Ritt06,Salg07}. Fig. 1 shows the band structure for such a Fourier-synthesized lattice for $V_1 = 5 E_r$ and $V_2 = 1.6 E_r$ used in our experiment, where $E_r=\hbar^2k^2/2m$ denotes the recoil energy, for two different values of the relative phase $\varphi$ between lattice harmonics. For the shown parameters, while the splitting between the first and the second excited Bloch band exhibits a nonzero value for $\varphi=0$, it vanishes for a phase of $\varphi=\pi$. The critical dependence on the relative phase between lattice harmonics is understood in terms of the splitting arising from both contributions of second order Bragg scattering of the usual lattice and of first order Bragg scattering of the higher spatial harmonic, where the contributions interfere constructively or destructively depending on $\varphi$ \cite{Salg07}. Of special interest is the case of destructive interference of the Bragg-scattering amplitudes, realized with $\varphi=\pi$ (see Fig. 1 right), for which at a suitable choice of lattice amplitudes the dispersion relation near the resulting crossing point becomes linear, i.e. relativistic, with an effective light speed  $c_{\scalebox{0.75}{\tn{eff}}}=2\hbar k/m \simeq 1.1$cm/s for the used 783nm laser wavelength. Both for vanishing and small splittings between the bands, we expect to be able to simulate relativistic physics, where a variation of the effective atomic Compton wavelength $\lambda_{\scalebox{0.75}{\tn{c,eff}}}=h/m_{\scalebox{0.75}{\tn{eff}}}c_{\scalebox{0.75}{\tn{eff}}} = 2 c_{\scalebox{0.75}{\tn{eff}}}h/\Delta E$, with $m_{\scalebox{0.75}{\tn{eff}}} = \Delta E/2c_{\scalebox{0.75}{\tn{eff}}}^2$ as the effective mass and $\Delta E$ as the size of the splitting, is possible by appropriate choice of amplitude and phase values of the lattice harmonics. In the limit of $\Delta E \rightarrow 0$ the effective Compton wavelength diverges. If we choose the zero point of the energy scale to be at the crossing, atomic population in the second excited band (above the crossing) corresponds to a particle-like excitation, in the below lying band of negative energy according to the St\"uckelberg-Feynman interpretation to a particle-like excitation propagating backwards in time, which is equivalent to a temporally forward propagating antiparticle excitation \cite{Bjork64}. Formally, the dynamics of atoms in the bichromatic lattice near the crossing between the first two excited bands can be described using a one-dimensional Dirac-like Hamiltonian (see \cite{Witt11} and Supplementary Material):
\begin{equation}\label{eq1}
H = m_{\scalebox{0.75}{\tn{eff}}}c_{\scalebox{0.75}{\tn{eff}}}^2\sigma_z + c_{\scalebox{0.75}{\tn{eff}}}\hat q\sigma_x + V_{\scalebox{.75}{\tn{slow}}}\rk{z},
\end{equation}
where $\sigma_x$ and $\sigma_z$ are Pauli matrices, $\hat q = -i\hbar \delta_z$ is the momentum operator and $V_{\scalebox{0.75}{\tn{slow}}}\rk{z}$ is an external potential varying much slower than the lattice periodicity. The two-component Hamiltonian acts on spinors $\psi = \rk{\psi_2,\psi_1}$, with $\psi_1$ and $\psi_2$ corresponding to course-grain atomic wave-functions in the upper and lower bands, respectively. Equation \ref{eq1} becomes exact in the limit $\abs{q}\ll \hbar k$ and $m_{\scalebox{0.75}{\tn{eff}}}c_{\scalebox{0.75}{\tn{eff}}}^2 \ll E_r$. One of the hallmark-effects of a relativistic dispersion is Klein-tunneling, which is a single-particle effect, so it can be equally well observed for bosons (as used in our experiment) and fermions \cite{Fesh58}.

\begin{figure}
\includegraphics[width=7.2cm]{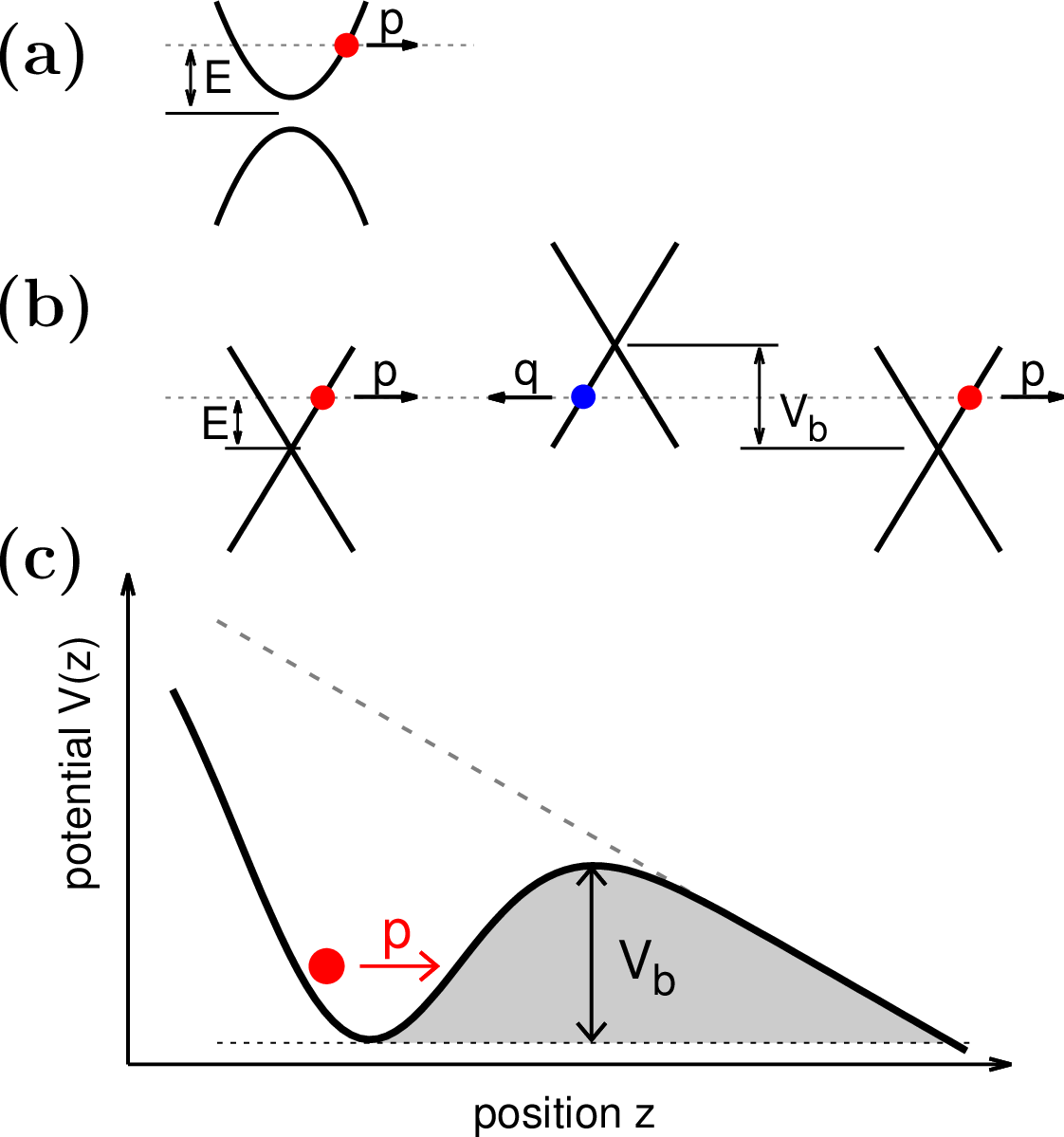}
\caption{Klein-tunneling of atoms through a potential barrier. (a) Relevant part of the dispersion relation in the lattice for a relative phase between lattice harmonics of $\varphi=0$, for which the large splitting suppresses a tunneling between bands. (b) For $\varphi=\pi$ the dispersion relation is linear and the two Bloch bands touch each other. Shown in the three diagrams is the variation of the atomic energy during passage of a potential barrier of height $V_b$. (c) Spatial distribution of the combined potential formed by gravitational and dipole trapping potential. Atoms can pass the potential well in the case of $\varphi=\pi$ due to their possibility to drop to below the band crossing in the Bloch spectrum when loosing potential energy, see the middle graph in (b).}
\end{figure}

In our experiment Klein-tunneling is investigated by monitoring the transmission through an external potential barrier that stands against the outcoupling of atoms from a far detuned optical dipole potential of depth $V_0$ by the earth's gravitational acceleration $g$. The spatial distribution of the combined potential $V_{\scalebox{0.75}{\tn{slow}}}\rk{z} = -V_0\exp\rk{-2\rk{z/\omega_0}^2}-mgz$ is shown in Fig. 2c, where the height $V_b$ of the potential barrier relatively to the minimum of the trapping potential can be adjusted by variation of the dipole trap beam power. The width of the potential barrier is of the same order as the used beam diameter $2\omega_0 \simeq\tn{46} \mu$m. For a typical atomic energy of one recoil energy below the maximum of the potential barrier, the estimated probability for usual nonrelativistic quantum tunneling is of the order $P_{\scalebox{0.75}{\tn{nr}}}=\exp\rk{-2\sqrt{2mE_r}z/\hbar}\simeq 10^{-170}$, i.e. completely negligible.

The situation however changes when quasirelativistic Klein-tunneling occurs, because the tunneling rate for this process does not decay exponentially with the width of the potential barrier. A quasirelativistic dispersion relation for ultracold rubidium atoms is induced using the Fourier-synthesized optical lattice, and Figs. 2a and 2b (left) indicate the relevant part of the atomic dispersion relation near the crossing between the first and the second excited Bloch-band for a relative phase of $\varphi=0$ and $\varphi=\pi$. The atoms are loaded at a quasimomentum $q$ well above the crossing region, but when proceeding towards the potential well on its rising edge loose kinetic energy, i.e. their momentum reduces. For a phase $\varphi=0$, the splitting between the first and second excited Bloch-band is comparatively large (see Fig.2a). This results in a small effective Compton wavelength, $\lambda_{\scalebox{0.75}{\tn{c,eff}}}\approx 6\mu$m, which is below the length of the rising edge of the barrier, and we expect no Landau-Zener tunneling into the lower band. The dispersion relation for atoms in the upper band then is Schr\"odinger-like. When the height of the potential barrier is larger than $q^2/2m_{\scalebox{0.75}{\tn{eff}}}$, the particle cannot pass through the barrier.

On the other hand, for a relative phase $\varphi=\pi$ between lattice harmonics, cf. Fig. 2b, the dispersion relation in the vicinity of the crossing point between the first and the second excited Bloch-band becomes ultrarelativistic. The effective Compton wavelength becomes larger than the widths of the edges of the barrier, and particles that approach the potential barrier and loose kinetic energy on the rising edge can be accelerated to below the crossing point between the second and the first excited Bloch-band (the “Dirac-point”), i.e. to states of negative energy of the energy scale shown in Fig.2c. Correspondingly, they can surpass higher potential barriers than in the “nonrelativistic” case of Fig. 2a. This corresponds to the case of Klein-tunneling. Our experiment simulates the conversion of a particle into a spatially backwards propagating antiparticle during the transmission of the barrier, whereafter the crossing point of bands again is passed on the tailing edge, so that particle-like states are again observed beyond the well. As in the case of the Klein-tunneling of electrons, this is equivalent to a double Landau-Zener passage between states of positive and negative energy respectively \cite{Bjork64,Kats06}.

The experiment proceeds by initially producing a Bose-Einstein condensate of rubidium atoms in an $m_F=0$ spin projection of the $F=1$ ground state within a dipole trapping potential formed by a focused CO$_2$-laser beam. To prepare atoms in the second excited Bloch-band, we initially leave the atoms in ballistic free fall by extinguishing the CO$_2$-laser dipole potential until the atoms have reached a momentum of 0.9$\hbar k$ by the earth’s gravitational force, and then apply a Doppler-sensitive Raman pulse transferring atoms to the $m_F=-1$ spin projection and imparting two photon recoil momentum, which increases the momentum to its desired value of $2.9\hbar k$. This initial momentum allows to load atoms into the second excited Bloch-band of the lattice potential (at the position $q=0.9 \hbar k$), and the CO$_2$-laser dipole potential required for the shaping of the desired slowly spatially varying potential barrier is again activated. The ballistic free fall during momentum preparation is about 2$\mu$m, i.e. well below the focal diameter of the trapping beam.

\begin{figure}
\includegraphics[width=8.5cm]{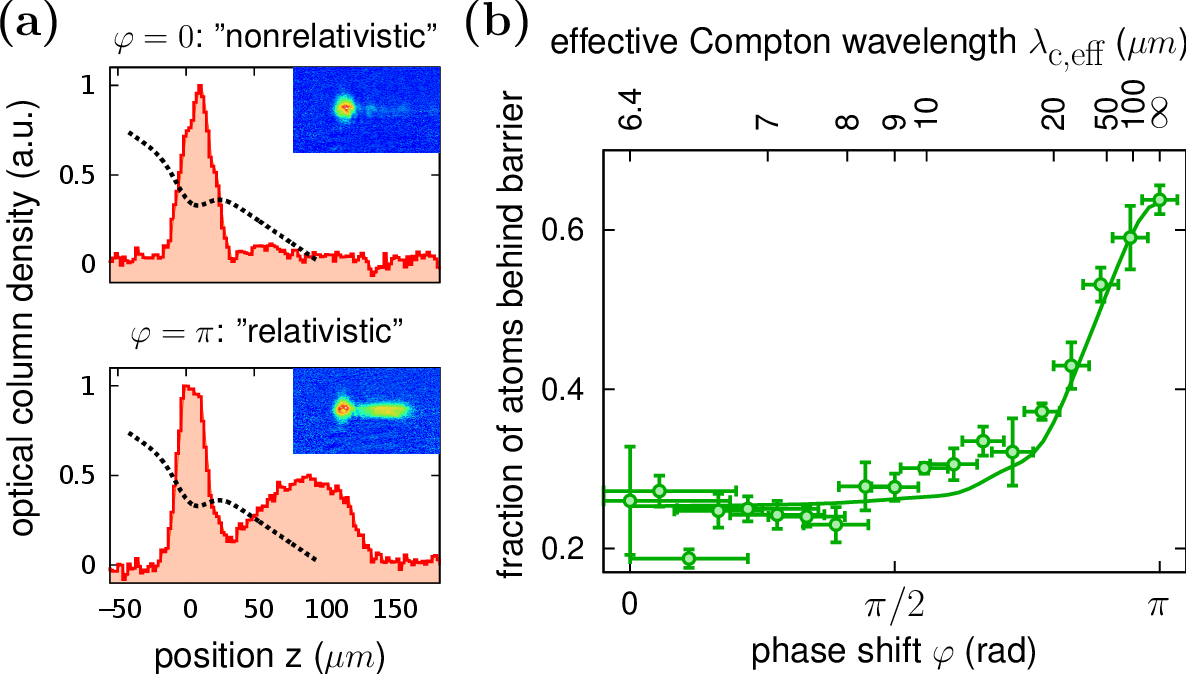}
\caption{(a) Cuts (solid red) through the measured spatial atomic distribution (insets) for a 5ms experiment time. Atoms proceeding from the center of the dipole trapping potential towards the potential barrier that detains against an outcoupling by the earth’s gravitational field for a relative phase between lattice harmonics of (top) $\varphi=0$ and (bottom) $\varphi=\pi$. The dotted black line indicates the potential $V_{\tn{\tiny slow}}\rk{z}$. (b) Relative atomic population beyond the potential barrier versus phase $\varphi$. The corresponding effective Compton wavelength $\lambda_{c,\tn{\tiny eff}}=2c_{\tn{\tiny eff}}h/\Delta E$ is shown on the top scale. The shown horizontal error bars refer to the latter scale and are dominated by the uncertainty in $\Delta E$, while the estimated uncertainty for the phase $\varphi$ (lower scale) is below the drawing size of the dots. The solid line is the result of a numerical integration of the relativistic wave equation (see eq. (\ref{eq1})). The only free fit parameters were amplitude and offset.}
\end{figure}

Fig.3a shows typical experimental data for the spatial atomic distribution, as recorded by absorption imaging at a time 5ms after preparation, for a relative phase $\varphi=0$ and $\varphi=\pi$ between lattice harmonics respectively. The height of the potential barrier was $V_b=5E_r$, while the average initial energy of the particle was $E= E_{\scalebox{.75}{\tn{kin}}}+m_{\scalebox{.75}{\tn{eff}}}c_{\scalebox{.75}{\tn{eff}}}^2=3.9 E_r$, with $m_{\scalebox{.75}{\tn{eff}}}c_{\scalebox{.75}{\tn{eff}}}^2$ of $0.5 E_r$ and $0$ for $\varphi=0$ and $\varphi=\pi$, respectively (see also Fig. 1a). The insets are false-colour shadow images of the atomic cloud, with the spot on the left-hand side corresponding to atoms near the trap location (i.e. before the barrier) and atoms on the right-hand side to particles that have transmitted the barrier. The solid red lines in the main images are cuts through the center, with the external potential indicated by dotted black lines. The data shows that for a relative phase of $\varphi=0$, almost all atoms remain in front of  the barrier, as expected. On the other hand, for $\varphi=\pi$, our experimental data show that most of the atomic population can be found beyond the potential barrier, equivalent to Klein-tunneling in this  optical lattice system. 

Note that for Klein-tunneling perfect transmission through the barrier is expected, when the energy splitting  $\Delta E$ would be zero. In our experiment, some 30\% of the population remains in front of the barrier, which we mainly attribute to atoms that did not take part in the Doppler-sensitive Raman transfer, and remain trapped in the CO$_2$-laser beam focus. We have measured the atomic population found beyond the barrier for variable values of the phase $\varphi$ between lattice harmonics, to investigate the case of intermediate values of the splitting $\Delta E$ between Bloch-bands. Corresponding data is displayed by the dots in Fig. 3b. The effective Compton wavelength becomes sufficiently large for Klein-tunneling only in a narrow region near $\varphi=\pi$ , corresponding to the ultrarelativistic case, while the atoms remain in front of the barrier for smaller phase values. The solid line is the result of a numerical simulation of the relativistic wave equation (see eq.1), in good agreement with the experimental data.

\begin{figure}
\includegraphics[width=7.9cm]{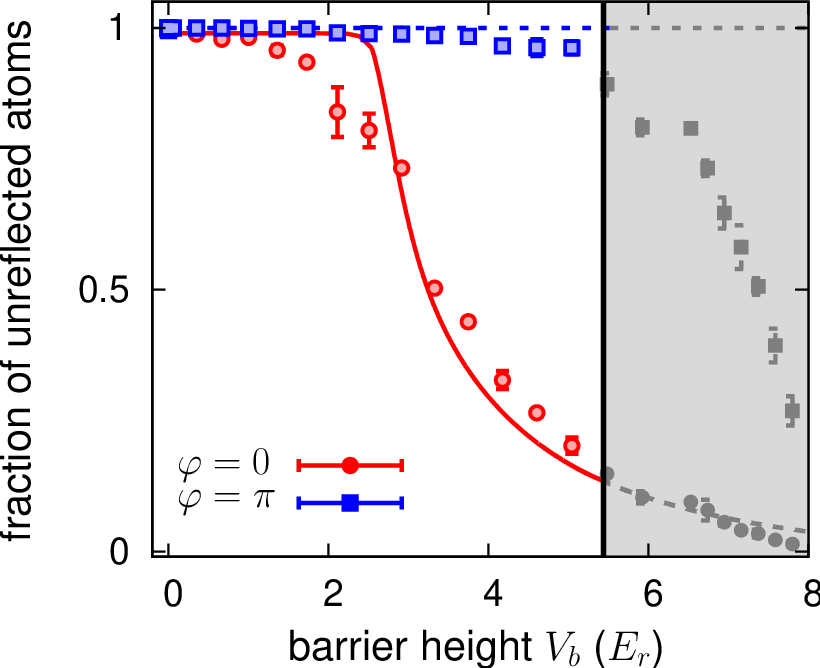}
\caption{Fraction of unreflected atoms versus the height of the potential barrier $V_b$ for a phase shift (solid red) $\varphi=0$ and (dashed blue) $\varphi=\pi$. The solid lines show numerical simulations. The grey shaded region corresponds to barrier height values where the atomic velocity reaches the edge of the Brillouin zone at $\abs{q}=\hbar k$. The desired dispersion relation is reached only in the white region. The experimental parameters are the same as in Fig. 3a except for the barrier height.}
\end{figure}

A striking prediction for Klein-tunneling is that the transmission through the barrier is expected to be independent of the barrier height, an issue in clear contrast to the expectations for nonrelativistic quantum mechanics. For a corresponding measurement in our system a momentum resolved time-of-flight measurement was employed to allow for a Stern-Gerlach separation of atoms that did not take part in the Raman transfer (see Supplementary Material). Fig.4 shows the relative signal of unreflected atoms, corresponding to Klein-tunneled atoms, versus the barrier height $V_b$. The grey shaded region corresponds to barrier height values where the atomic velocity reaches the edge of the lattice Brillouin zone at $\abs{q}=\hbar k$, for which the bottom of the first exited band can be reached, i.e. only in the left white region the desired dispersion relation is achieved, with larger accuracy when remaining far from the shaded region. The blue squares are data recorded for $\varphi=\pi$, corresponding to a quasirelativistic dispersion, for which this signal remains at a high value within nearly the complete white region, illustrating the prediction of Klein-tunneling being independent of the barrier height with good accuracy. On the other hand, a pronounced loss of this signal at large barrier heights is observed for  $\varphi=0$ (red dots), corresponding to a nonrelativistic dispersion. Quantum tunnelling remains negligible due to the large spatial width of the barrier. The finite width of the kinetic atomic energy distribution here softens the otherwise expected sharp decay for $E_{\scalebox{0.75}{\tn{kin}}} < V_b$. The shown solid and dashed lines are the result of a numerical simulation, which for the “relativistic” case are in good agreement with the data and also qualitatively reproduce the “nonrelativistic” case.

To conclude, we report an experiment demonstrating the analog of Klein-tunneling, as a proof-of-principle experiment testing relativistic wave equation predictions, with ultracold atoms in a bichromatic optical lattice. By tuning the relative phase between lattice harmonics the atomic dispersion can be tuned continuously from the nonrelativistic to the ultrarelativistic case, though the atoms move at a velocity 10 orders of magnitude below the speed of light.

For the future, we expect that ultracold atoms in optical lattices allow quantum simulations of a wide range of effects of both linear and nonlinear Dirac-dynamics. Perspectives include the verification of theoretical high energy physics predictions \cite{Merkl10}, as chiral confinement, and other results of massive Thirring models \cite{Cirac10,Chang75,Lee75}.

\begin{acknowledgments}
Financial support of the DFG is acknowledged. We thank A. Rosch, L. Santos, 
D. Witthaut, H. Kroha, and K. Ziegler for discussions.
\end{acknowledgments}

\bibliographystyle{apsrev}

\newpage
\onecolumngrid
\appendix

\begin{center}
{\bf Supplementary Material}\\[.5cm]
\end{center}

\section*{\it Experimental Approach and Setup}

The experiment is based on a Bose-Einstein condensate of rubidium atoms ($^{87}$Rb) in a Fourier-synthesized optical lattice used to taylor the atomic dispersion relation. Rubidium atoms are cooled to quantum degeneracy evaporatively in an optical dipole trap realized by focusing a beam of mid-infrared radiation derived from a CO$_2$-laser with wavelength near 10.6 $\mu$m. During the final stages of evaporation, a magnetic gradient field is switched on, resulting in a spin-polarized Bose-Einstein condensate with about $6\cdot 10^4$ atoms in the $m_F=0$ Zeeman sublevel of the $F=1$ hyperfine ground state component [S1].

\begin{figure}[h]
\includegraphics[width=8cm]{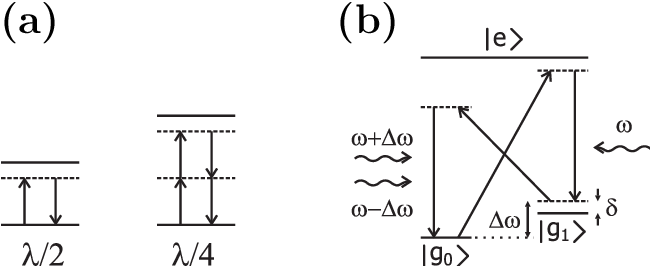}
\renewcommand{\figurename}{Fig. S}
\caption{(a) Left: Two-photon processes in a conventional optical standing wave lattice, which are responsible for a periodic potential of $\lambda/2$ spatial periodicity. Right: Four-photon processes yielding a contribution to the lattice potential of $\lambda/4$ spatial periodicity. However, unwanted second order contributions that yield a usual standing wave lattice potential, dominate in this simple approach. (b) Improved scheme for generating a periodic potential with $\lambda/4$ spatial periodicity based on a three-level atom driven by an optical field of frequency $\omega$ and two counterpropagating optical fields of frequencies $\omega-\Delta\omega$ and $\omega+\Delta\omega$.}
\label{SOM}
\end{figure}

The Fourier-synthesized optical lattice potential has contributions of two lattice harmonics with spatial periodicities $\lambda/2$ and $\lambda/4$ respectively, where $\lambda$ denotes the wavelength of the driving laser, tuned 3 nm to the red of the rubidium D2-line. For the fundamental spatial frequency of $\lambda/2$ periodicity, a conventional optical standing wave potential is used, achieved with counterpropagating laser beams of same optical frequency. The atoms here undergo virtual two-photon processes of absorption of one photon from one beam and stimulated emission of another photon into a counterpropagating beam. A potential with spatial periodicity $\lambda/4$ could in principle be achieved by replacing each of the absorption and emission cycles with a stimulated process induced by two photons, as indicated in Fig. S\ref{SOM} (a). The scheme used in our experiment to generate a lattice with $\lambda/4$ periodicity, is shown in Fig. S\ref{SOM}b [S2,S3]. In this improved approach, absorption (stimulated emission) processes have been exchanged by stimulated emission (absorption) processes of an oppositely directed photon. The high resolution of Raman transitions between two stable ground states $\ket{g_-}$ and $\ket{g_+}$ over an excited state $\ket{e}$ here allows to clearly separate the desired four-photon process from lower order contributions in frequency space. A set of three different optical frequencies is used to drive the transitions, with two copropagation laser fields of frequencies $\omega-\Delta\omega$ and $\omega+\Delta\omega$ respectively and a counterpropagating beam with frequency $\omega$.

We use the rubidium $F = 1$ electronic ground state Zeeman components $m_F = -1$ and $0$ as levels $\ket{g_-}$ and $\ket{g_+}$, and the $5\tn{P}_{3/2}$ excited state manifold as state $\ket{e}$. A magnetic bias field of 1.8 G removes the degeneracy of the Zeeman sublevels. By combining lattice potentials of $\lambda/2$ and $\lambda/4$ spatial periodicities, a variably shaped optical lattice can be Fourier-synthesized. The light to generate the lattice potential is delivered by a high power diode laser system. The emitted near-infrared beam is split into two, from which the two counterpropagating beams for generation of the periodic atom potentials are derived. In each of the beams, an acoustooptic modulator generates all required optical frequencies in the corresponding beam path. We estimate that the accuracy with which the size of the energetic splitting $\Delta E$ between the first two excited Bloch bands of the variably shaped optical lattice can be controlled is roughly $E_r/20$.

To prepare atoms in the desired position of the Bloch-spectrum after creation of the Bose-Einstein condensate the atomic cloud is left in free fall by deactivating the CO$_2$-laser trapping beam for about 0.5 ms, after which the atoms have accelerated to a momentum of 0.9$\hbar k$ due to the earth’s gravitational field, and then apply a Doppler-sensitive Raman pulse of 40 $\mu$s length transferring atoms from the $m_F=0$ to the $m_F=-1$ Zeeman sublevel and imparting another two photon recoil momentum. This increases the atomic momentum to 2.9$\hbar k$, which completes the velocity preparation sequence. The Raman transfer selects a typical Doppler width of $\pm 0.3 \hbar k/m$ width from the initial atomic distribution obtained after expansion of the interacting condensate atoms. Atoms are then loaded adiabatically into the second excited Bloch-band of the Fourier-synthesized optical lattice potential. Further, the far detuned CO$_2$-laser beam required for the shaping of the desired slowly varying potential
\begin{equation}\label{eq-V}
V_{\scalebox{0.75}{\tn{slow}}}\rk{z}=-V_0\exp\rk{-2\rk{z/\omega_0}^2}-mgz
\end{equation}
is reactivated, which imposes a potential barrier for the vertically downwards propagating atoms.

\section*{\it Spatial and momentum-resolved measurements of ultracold atomic ensembles}

During the course of the experiment, two types of measurements were used for an analysis of the atomic final state. In a first series of experiments (see Figs. 3(a) and 3(b) for results), the spatial atomic distribution was recorded by means of an absorption image recorded after a 5ms long experiment time of both CO$_2$-laser dipole potential and the Fourier-synthesized lattice potential interacting with the atoms. Given the known geometry of the slowly varying potential $V_{\scalebox{0.75}{\tn{slow}}}\rk{z}$ (which is the sum of contributions from the CO$_2$-laser dipole trapping potential and of the gravitational field, see eq. \ref{eq-V}), we can from their different spatial locations clearly distinguish between atoms, which remain near the focus of the CO$_2$-laser beam (i.e. in front of the barrier) and atoms that have been outcoupled. However, residual $m_F=0$ atoms which did not take part in the Raman acceleration transfer also remain trapped near the CO$_2$-laser beam focus, and in this detection method contribute to lower value of the observed transmission.

For the data shown in Fig.4 a second detection method, based on a far field time-of-flight technique that allows for a Stern-Gerlach separation of the individual Zeeman sublevels, was used. After a typically 2 ms long experimental cycle the optical lattice beams were switched off, and a magnetic field gradient was activated. An absorption image was then recorded after a 10 ms long ballistic time-of-flight period, during which atoms in different Zeeman sublevels spatially separate from each other by the Stern-Gerlach force, so that residual population in $m_F=0$ can be distinguished from atoms in $m_F=1$. Due to the wave nature of the atomic Bose-Einstein condensate, atomic clouds are formed in different diffraction orders during the time-of-flight. Atoms that have undergone the analog of Klein-tunneling are expected to be in the same diffraction order as prior to the experiment (this is the signal shown on the vertical axis of Fig. 4). An analysis of the atomic diffraction order allows us to distinguish tunneled from untunneled atoms, with the latter being reflected by the potential barrier.

The shown theory curves in Fig. 3b and Fig. 4 have been obtained by numerical integration of the one-dimensional time-dependent relativistic wave equation (eq. 4). The used initial conditions were an initial mean momentum $q=0.9\hbar k$ and a Gaussian wavepacket size of $\sigma=4.5 \mu$m FWHM spatial width (which yields an associated uncertainly limited width of the momentum distribution). For the case of the spatially resolved measurements of Fig. 3b, amplitude and offset were used as free fit parameters to account for residual atoms in the $m_F=0$ Zeeman sublevel remaining before the barrier and uncertainties in the barrier height respectively. For the momentum resolved technique used in the measurement with results shown in Fig.4, the Zeeman sublevels can be clearly resolved. For the theory curve shown in this figure for the case of $\varphi=0$, only the positive energy branch is relevant in the calculation, and the visible width of the dropoff is determined by the finite spread of the atomic velocity distribution. Quantum tunnelling here remains neglibible due to the large width of the barrier, even if one accounts for a modification of the effective mass in the lattice (compared to the free atomic mass, the effective mass in the lattice can lighter by up to an order of magnitude for typical experimental parameters).  A residual discrepancy between theoretical and experimental curves is here mainly attributed to partly overlapping atomic clouds in the far field image of reflected and transmitted atoms. Within the 2 ms experiment time, transmitted atoms (i.e. atoms outcoupled from the dipole potential) during their ballistic free fall already reach the end of the Brillouin zone and perform Bloch oscillations. In the far field image, the corresponding atomic cloud then partly overlaps with the signal of atoms reflected by the barrier. The corresponding overlap introduces systematic uncertainties in this far field time-of-flight technique. Other (smaller) effects, which affect the theory curves in both of the figures, include the nonlinearity if the dispersion for larger values of $q$.

\section*{\it Effective relativistic wave equation}

The adiabatic potential created with the bichromatic lattice is of the form $V\rk{z}=V_1/2\cos\rk{2kz} + V_2/2\cos\rk{4kz+\varphi}$, where $V_1$ and $V_2$ are the potential depths of the lattice potentials with spatial periodicities $\lambda/2$ and $\lambda/4$ respectively, and $\varphi$ denotes the relative phase between spatial harmonics. We here show that the equation of motion in the periodic potential can be described by an effective relativistic wave equation for particles near the center of the Brillouin zone in the first two excited Bloch-bands. Atoms in the bichromatic lattice potential can be described by the Hamiltonian
$$H_{0,SG} = \frac{\hat p^2}{2m} + V\rk{z},$$
with $\hat p$ as the momentum operator. Solutions to the stationary one-dimensional Schr\"odinger equation with this Hamiltonian can be found in terms of a Bloch ansatz as a product of a plane wave  and a function  with the same periodicity as the lattice potential as $\phi_q^n\rk{z} = \exp\rk{iqz/\hbar}\cdot u_q^n\rk{z}$, where the index $n$ is an integer number and indexes the corresponding Bloch band and $q$ is the quasimomentum that is conventionally restricted to the reduced Brioullin zone $q=\e{-\hbar k, \hbar k}$ [S4]. We obtain an eigenvalue equation
\begin{equation}
H_{0,SG}\phi_q^n\rk{z} = E_q^n\phi_q^n\rk{z}.
\end{equation}
Since both the periodic potential and the function have the same periodicity as the lattice, they can be written as a Fourier series. The Bloch wavefunction can be written in the form
$$
\phi_q^n\rk{z} = \sum_s c_s \exp\rk{i\rk{q+2s\hbar k}z/\hbar}
$$
which is inserted into the stationary Schr\"odinger equation (1). The obtained eigenvalue equation is
\begin{equation}
\sum_{s'}M_{s,s'}c_{s'} = E_q^nc_s
\end{equation}
where $M_{s,s'}$ can be written in the form
$$M = \left(\begin{array}{ccc}\frac{1}{2m}\rk{q-2\hbar k}^2  & \frac {V_1}{4}  & \frac{V_2}{4}e^{i\varphi} \vspace{0.3cm} \\ \vspace{0.3cm} \frac {V_1}{4} & \frac{1}{2m} q^2 & \frac {V_1}{4} \\ \frac{V_2}{4}e^{-i\varphi} & \frac {V_1}{4} & \frac{1}{2m}\rk{q+2\hbar k}^2 \end{array}\right),$$
when we restrict ourselves to the ground and the first two excited Bloch bands. More specifically, we are here interested in the region around the band splitting between the first and the second excited band. In previous work it has been demonstrated that the size of this splitting depends critically on the relative phase (and amplitudes) of the lattice harmonics [S2]. Assuming that the quasimomentum $\abs{q}\ll \hbar k$, i.e. we consider only the region close to the avoided crossing, adiabatic elimination of the ground band leads to a reduced 2x2 matrix
\begin{equation}
M_{red} = \left(\begin{array}{cc}-\frac{2\hbar k}{m}q & \Delta E/2 \vspace{0.3cm} \\ \Delta E/2 & \frac{2\hbar k}{m}q  \end{array}\right)= \left(\begin{array}{cc}-qc_{\scalebox{0.75}{\tn{eff}}} & m_{\scalebox{0.75}{\tn{eff}}}c_{\scalebox{0.75}{\tn{eff}}}^2 \vspace{0.3cm} \\ m_{\scalebox{0.75}{\tn{eff}}}c_{\scalebox{0.75}{\tn{eff}}}^2 &qc_{\scalebox{0.75}{\tn{eff}}}  \end{array}\right),
\end{equation}
with the basis states $\{\exp\rk{i\rk{q-2\hbar k}z/\hbar}\ket{q-2\hbar k},$ $\exp\rk{i\rk{q+2\hbar k}z/\hbar}\ket{q+2\hbar k}\}$, where we are only left with the first two excited Bloch bands. The eigenenergies here have been measured relatively to the position of the band crossing. Further, we have used  as the effective light speed and an effective mass $m_{\scalebox{0.75}{\tn{eff}}}=\frac{\Delta E}{2c_{\scalebox{0.75}{\tn{eff}}}^2} = \frac{1}{c_{\scalebox{0.75}{\tn{eff}}}^2}\cdot \abs{\frac{V_1^2}{64E_r}+\frac{V_2}{4}e^{i\varphi}}$, with the latter depending on the amplitudes of the potential depths of the lattice harmonics and on the relative phase. Applying a rotation to eq. (3) with a unitary transformation matrix
$$1/\sqrt{2}\left(\begin{array}{cc}1 & 1 \\ -1 & 1\end{array}\right)$$
and replacing $q$ by the corresponding operator finally gives an effective Hamiltonian
$$H_0 = m_{\scalebox{0.75}{\tn{eff}}}c_{\scalebox{0.75}{\tn{eff}}}^2\sigma_z + c_{\scalebox{0.75}{\tn{eff}}}\hat q \sigma_x$$
For a derivation of the full effectively relativistic wave equation Hamiltonian $H = m_{\scalebox{0.75}{\tn{eff}}}c_{\scalebox{0.75}{\tn{eff}}}^2\sigma_z + \hat q c_{\scalebox{0.75}{\tn{eff}}}\sigma_x + V_{\scalebox{0.75}{\tn{slow}}}\rk{z}$  with an additional external potential $V_{\scalebox{0.75}{\tn{slow}}}\rk{z}$ (see eq. \ref{eq-V}), which acts on spinors $\psi = \rk{\psi_1,\psi_2}$ with $\psi_1$ and $\psi_2$ corresponding to course-grain atomic wave-functions in the upper and lower bands respectively, see Ref. [S5]. This derivation assumes that the external potential varies slowly on the scale of the lattice periodicity. The corresponding time-dependent wave equation
\begin{equation}
i\hbar \frac{\partial}{\partial t}\psi = H \psi
\end{equation}
has a Dirac-like form when using the described Hamiltonian. By forming the temporal derivative of one of the spinor wavefunction components and reinserting the two components into the obtained equation, this can readily be shown to be equivalent to the one-dimensional Klein-Gordon equation
\begin{equation}
\rk{i\hbar \frac{\partial}{\partial t} - V_{\scalebox{0.75}{\tn{slow}}}\rk{z}}^2\Phi = \rk{m_{\scalebox{0.75}{\tn{eff}}}^2c_{\scalebox{0.75}{\tn{eff}}}^4 + \rk{\frac{\hbar}{i}\frac{\partial}{\partial z}}^2c_{\scalebox{0.75}{\tn{eff}}}^2}\Phi
\end{equation}
where $\Phi$ is a (scalar) course-grain Klein-Gordon wavefunction. Eq. (5) is the well-known relativistic wave equation for a bosonic particle [S6].

\vspace{2cm}
[S1] G. Cennini, G. Ritt, C. Geckeler, M. Weitz
Phys. Rev. Lett {\bf 91}, 240408 (2003).

[S2]
T. Salger, C. Geckeler, S. Kling, M. Weitz
Phys. Rev. Lett {\bf 99}, 190405 (2007).

[S3]
P.R. Berman, B. Dubetsky, J.L. Cohen,
Phys. Rev. A {\bf 58}, 4801 (1998).

[S4]
See e.g.: N.W. Ashcroft, N.D. Mermin
{\it Solid State Physics} (Saunders College Publishing, New York, 1976)

[S5]
D. Witthaut et al.,
Phys. Rev. A (in press).

[S6]
See e.g.: G. Baym,
{\it Lectures on Quantum Mechanics} (Westview Press, Boulder, 1974)

\end{document}